# Supervised-learning-based Optimal Thermal Management in an Electric Vehicle

Youngjin Kim,

**Pohang University of Science and Technology.**

**ABSTRACT** Due to the increasing market share of electric vehicles (EVs), the optimal thermal management (TM) of batteries has recently received significant attention. Optimal battery temperature control is challenging, requiring a detailed model and numerous parameters of the TM system, which includes fans, pumps, compressors, and heat exchangers. This paper proposes a supervised learning strategy for the optimal operation of the TM system in an EV. Specifically, for TM subsystems, individual artificial neural networks (ANNs) are implemented and trained with data obtained under normal EV driving conditions. The ANNs are then interconnected based on the physical configuration of the TM system. The trained ANNs are replicated using piecewise linear equations, which can be explicitly integrated into an optimization problem for optimal TM scheduling. This approach enables the application of a mixed-integer linear programming solver to the problem, ensuring the global optimality of the solution. Simulation case studies are performed for the two operating modes of the TM system: i.e., integrated and separate modes. The case study results demonstrate that the ANN-based model successfully reflects the operating characteristics of the TM system, enabling accurate battery temperature estimation. The proposed optimal TM strategy using the ANN-based model is verified to be effective in reducing the total energy consumption, while maintaining the battery temperature within an acceptable range.

## NOMENCLATURE

*Acronyms*:

| | |
|---|---|
| AC | air conditioning |
| ANN | artificial neural network |
| EV | electric vehicle |
| EWP | electric water pump |
| LTR | low-temperature radiator |
| MILP | mixed integer linear programming |
| MPC | model predictive control |
| NARX | nonlinear auto-regressive network with exogenous inputs |
| NMSE | normalized mean squared error |
| PE | power-electronics |
| ReLU | rectified linear unit |
| SL | supervised learning |
| TM | thermal management |

*Sets and Indices*:

| | |
|---|---|
| $t, \tau$ | superscripts for time |
| $min, max$ | subscripts for minimum and maximum values |
| $i, j, o, h$ | subscripts for $i_{th}$ and $j_{th}$ neurons, output neuron, and $h_{th}$ hidden layer |
| $s, s'$ | subscripts for $s_{th}$ and $s'_{th}$ subsystems |
| $se$ | subscripts for segments of linearized activation function |
| $u$ | subscripts for controllable devices in TM system |
| $b, cp, fn, p, pe$ | subscripts for battery, compressor, radiator fan, electric water pumps, and power electronics |
| $bi, ro$ | subscripts for battery inlet and radiator outlet |
| $c, e, y$ | subscripts for controllable inputs, environmental inputs, and outputs |
| $pre$ | subscripts for data during a previous scheduling period |
| $k$ | subscripts for $k_{th}$ data in training dataset |
| $ab, vs, f$ | subscripts for ambient temperature, vehicle speed, and coolant flow rate |
| $avg, avgs$ | subscripts for average value and average of standard deviations |
| $\pi$ | subscripts for $\pi_{th}$ combination of hyper-parameters |
| $\varepsilon$ | subscripts for $\varepsilon_{th}$ incremental variation in selected inputs |

*Parameters*:

| | |
|---|---|
| $N_t$ | number of scheduling time |
| $N_h, N_i, N_j$ | numbers of hidden layers and neurons |
| $N_{se}$ | number of segments for linearization |
| $N_c$ | number of controllable devices |
| $N_Y$ | number of the training datasets of **Y** with respect to time |
| $IW_{ij}$ | weighting coefficient for connection from $i_{th}$ input neuron to $j_{th}$ hidden neuron |
| $HW_{ij}$ | weighting coefficient for connection from $i_{th}$ hidden neuron to $j_{th}$ hidden neuron |
| $LW_j$ | weighting coefficient for connection from $j_{th}$ hidden neuron to the output neuron |
| $b_{j,h,s}, b'_{j,h,s}$ | biases for $j_{th}$ neuron in $h_{th}$ hidden layer in $s_{th}$ subsystem |
| $b_{o,s}$ | biases for output layer in $s_{th}$ subsystem |
| $d_c, d_e, d_y$ | maximum time step delays in controllable inputs, environmental inputs, and outputs |
| $l_{se}$ | slope of linearized activation function for $se_{th}$ segment |
| $r_0, r_1, r_2$ | constants for boundaries of linear segments |
| $e_{NMSE}$ | normalized mean squared error |
| $Q_{\pi,\varepsilon}{}^t$ | evaluation index of ANN for $\varepsilon_{th}$ incremental input variation at time $t$, given $\pi_{th}$ hyper-parameter combination |
| $Q_{avg,\pi,\varepsilon}$, $Q_{avgs,\pi,\varepsilon}$ | average values and average of standard deviations of $Q_{\pi,\varepsilon}{}^t$ |
| $c_P, c_T$ | weighting coefficients on the costs for power consumption and battery temperature variation |
| $E_{ec}$ | total power (energy) consumption of TM system |
| $V_{vs}, t_{dr}$ | vehicle speed and driving time |
| $I_b, AC$ | battery current and AC switch status |
| $T_{ab}, T_{pe}$ | ambient and PE temperatures |

*Variables*:

| | |
|---|---|
| $T_b, T_{bi}, T_{ro}$ | temperatures of battery, battery-inlet coolant, and radiator-outlet coolant |
| $P_p, P_{fn}, P_{cp}$ | power inputs of EWP, radiator fan, and compressor |
| $P_u{}^t$ | power inputs of each controllable device $u$ at time $t$ |
| $m_f$ | coolant flow rate |
| $\mathbf{X_C, X_E, X_F, Y}$ | controllable inputs, environmental inputs, feedback inputs, and outputs of NARXs |
| $X_{i,s}{}^t, x_{i,s}{}^t$ | $i_{th}$ input and normalized input of $s_{th}$ ANN at time $t$ |

| $Y_s^t, y_s^t$ | output and normalized output of $s_{th}$ ANN at time $t$ |
| $n_{j,h,s}^t, m_{j,h,s}^t$ | input and output of activation function for $j_{th}$ neuron in $h_{th}$ hidden layer of $s_{th}$ ANN at time $t$ |
| $q_{se,j,h,s}^t, v_{j,h,s}^t$ | continuous and binary variables for $se_{th}$ linear segment of activation function input for $j_{th}$ neuron in $h_{th}$ hidden layer of $s_{th}$ ANN at time $t$ |
| $X_{\pi,\varepsilon}^t, Y_{\pi,\varepsilon}^t$ | input and output data with incremental variation $\varepsilon \cdot \Delta X$ |

## I. INTRODUCTION

The market share of electrified vehicles has been increasing continuously due to environmentally friendly policies and the falling prices of battery packs [1], [2]. The supply of electric vehicles (EVs) has significantly increased, along with those of other types of vehicles, such as plug-in hybrid electric vehicles (PHEVs), hybrid electric vehicles (HEVs), and fuel cell hybrid vehicles (FCHVs) [3], [4]. As large battery capacity is required for EV driving, significant attention has been paid to EV thermal management [5]–[7] for battery lifetime and performance improvement.

As shown in Table 1, many studies were conducted on optimal thermal management (TM) of batteries, involving physics-based modeling of TM systems for various types of vehicles. However, physics-based model predictive control (MPC) often requires numerous modeling parameters to reflect nonlinear operating characteristics of TM system components such as batteries, pumps, compressors, and heat exchangers. Most of the modeling parameters are unknown and time-varying, degrading the performance and reliability of physics-based MPC in practice.

The aforementioned issue can be resolved using big data and machine learning. The data on vehicle driving conditions and TM system operations are becoming increasingly available, mainly due to developments in internet-of-things technologies [8]. With these data, artificial neural networks (ANNs) can be trained via supervised learning (SL) to model the characteristics of TM system operations and corresponding battery temperature variations under various driving conditions. In other words, the trained ANNs can reflect the variations in battery temperature with changes in TM system operation, for example, with respect to the fan power, coolant flow rate, vehicle speed, and ambient temperature.

The trained ANNs can be used for both thermal load prediction and optimal TM system operation, as discussed in [9] and [10], although these studies were focused on buildings, rather than EVs. For optimal operation, an optimization problem is formulated using the trained ANNs, relieving the necessity of acquiring the physics-based modeling parameters of the TM system. Therefore, compared to physics-based MPC, ANN-based MPC can be readily applied to various sizes and types of TM systems that are either implemented for field tests or used daily in practice.

In [9] and [10], the optimization problems for optimal TM system operation were solved using various nonlinear algorithms such as sequential quadratic programming (SQP), dynamic programming (DP), and Pontryagin's maximum principle (PMP). However, such nonlinear solvers cannot guarantee the global optimality of the solution within reasonable computational time. Simple approaches were also applied, considering practical applicability, where a pre-determined operating rule was implemented [11] and an optimum was searched by iteratively investigating the combinations of controllable, discretized inputs [12]. In [13]–[16], the objective function consisted of the terms related to the total power consumption and corresponding battery temperature variation. The objective function was also established considering the heat loss [11], [17] and the costs for fuel consumption and gas emission [18]. For optimal TM, the power inputs of fans and pumps were commonly considered as controllable inputs mainly due to the simplicity of device modeling. Compressors are rather complicated but have high power ratings, significantly affecting the power consumption and battery temperature. In [17] and [18], the output currents of the battery were actively controlled, for example, to reduce the power and heat losses for the case in which the EVs were parked in cold weather.

This paper proposes a new SL-based strategy for the optimal operation of the TM system in an EV. Specifically, in this strategy, an ANN-based model is implemented for each subsystem of the TM system that affects the battery temperature. The individual ANNs are trained with data obtained under normal EV driving conditions and then interconnected, based on the physical relationships between the subsystems. The trained ANNs are then represented using a set of explicit, piecewise linear equations, which can be directly integrated into the constraints of an optimization problem for the optimal TM. This approach enables the application of mixed-integer linear programming (MILP) to the optimization problem, ensuring the global optimality of the solution within reasonable computational time. Case studies are performed for common driving cycles: i.e., urban dynamometer driving schedule (UDDS) and highway fuel economy test (HWFET) schedule. The results confirm that the interconnected ANN successfully reflects the operating characteristics of the TM system, enabling accurate battery temperature estimation. Consequently, the proposed strategy using the ANN-based model effectively reduces the total power consumption of the TM system while maintaining the EV battery temperature within an acceptable range.

The main contributions of this paper are summarized as follows:
• To our best knowledge, this is the first study in which an ANN-based model of a TM system has been implemented and replicated using piecewise linear equations for integration into the optimization problem. This mitigates the necessity of obtaining physics-based modeling parameters and hence improves the applicability to various types of EVs.
• For each TM subsystem, individual ANNs are trained and interconnected based on the physical relationships between the subsystems. This process enhances the modeling performances of the ANNs, improving the accuracy of the battery temperature estimation.

TABLE 1. Previous studies of optimal TM system operation for electrified vehicles

| Ref. | Vehicle type | Modeling | Optimization algorithm[1] | Objective function[2] | | | Controllable devices | | | | |
|---|---|---|---|---|---|---|---|---|---|---|---|
| | | | | power consumption | battery temperature | others | pump | fan | compressor | valve | battery |
| Proposed | EV | Interconnected ANN | MILP | ○ | ○ | | ○ | ○ | ○ | | |
| [11] | EV | Physics-based | RB | | | SHD | ○ | ○ | | ○ | |
| [12] | HEV | | - | | ○ | | ○ | ○ | | | |
| [13] | PHEV | | SQP | ○ | ○ | | ○ | | ○ | ○ | |
| [14] | EV | | PMP | ○ | ○ | | ○ | | | | |
| [15] | PHEV | | SQP | ○ | ○ | | ○ | | ○ | ○ | |
| [16] | HEV | | DP | ○ | ○ | | ○ | ○ | | | |
| [17] | HEV, EV | | DP | ○ | | PL, HL | | | | | ○ |
| [18] | PHEV | | SQP | | | CF, CG | | | | | ○ |

[1]: SQP: sequential quadratic programming, PMP: Pontryagin's maximum principle, DP: dynamic programming, RB: rule-based
[2]: PL: power loss, HL: heat loss, SHD: specific heat dissipation, CF: cost for fuel consumption, CG: cost for gas emission

• A simple procedure has been developed to select ANN architectures for minimal over-fitting. This enhances the generalization capability of the ANNs in reflecting the operating characteristics of the TM subsystems, improving the performance of the SL-based TM system operation.

• Case studies have been performed for two TM modes: i.e., integrated and separate modes. The case study results confirm the effectiveness of the proposed strategy for the two TM modes and two common driving cycles in reducing the total power consumption while maintaining the battery temperature within an acceptable range.

The remainder of this paper is structured as follows: Section II presents the ANN-based model of the TM system. Section III explains the ANN model linearization and the optimization problem. Section IV presents the case study results. Section V concludes the paper.

## II. ANN-BASED THERMAL MANAGEMENT
### A. THERMAL MANAGEMENT SYSTEM

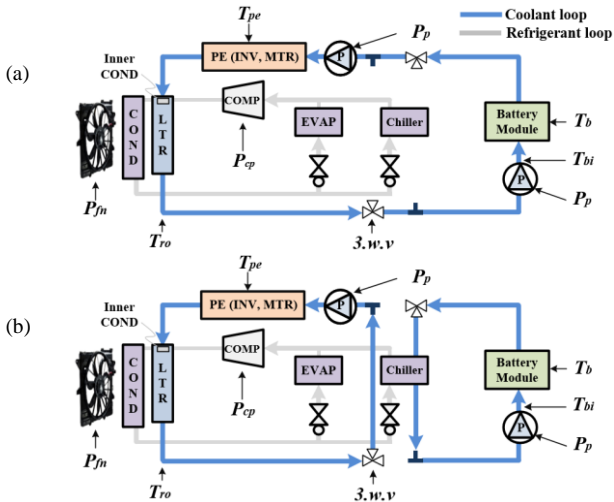

FIGURE 1. Schematic diagrams of a TM system in an EV: (a) integrated and (b) separate modes.

TABLE 2. Two operating modes of the TM system

| TM system operation | Integrated mode | Separate mode |
|---|---|---|
| Levels of $T_{ab}$, $T_{bi}$, and $T_b$ | low | high |
| Initiating conditions | $T_{bi} \leq T_b$ | $T_b < T_{ro}$ |
| Devices for $T_b$ control | EWPs, fan | EWP, fan, compressor |

Fig. 1(a) and (b) show the schematic diagrams of a common TM system in an EV [5] for the integrated and separate modes, respectively. For battery temperature control, the TM system includes three main types of controllable devices: i.e., a compressor, a radiator fan, and two electronic water pumps (EWPs) [19]. It also contains circulation loops via which the coolant delivers and exchanges heat with the ambient air and the refrigerant in the integrated and separate modes, respectively. As shown in Table 2, the three-way valves operate based on the relative differences between the battery temperature and coolant temperatures at the battery inlet and radiator outlet, resulting in two coolant loop pathways for the integrated and separate modes.

The integrated mode is initiated when $T_b$ is maintained within a stable range and $T_{bi}$ is sufficiently low to cool down the battery (i.e., $T_{bi} \leq T_b$). Specifically, the EWPs operate with the same speed, so that the coolant flows into the battery module and then the power-electronics (PE) module with consistent $m_f$. At the low-temperature radiator (LTR), the coolant exchanges the heat collected from the battery and PE modules with the ambient air. The coolant with low $T_{ro}$ then flows into the battery again. The operation of the radiator fan improves the rate of heat exchange between the coolant and ambient air, further reducing $T_{ro}$. On the other hand, the separate mode starts when $T_{ro}$ is higher than $T_b$, for example, due to the high ambient temperature. The coolant loop is then divided into two, as shown in Fig. 1(b). Consequently, $T_b$ control is achieved by delivering the heat collected from the battery to the refrigerant at the chiller and then to the ambient air at the condenser. The compressor can operate not only to achieve air conditioning (AC) inside the vehicle but also to improve the efficiency of battery heat delivery [5]. This leads to the considerable reduction of $T_b$ at the expense of an increase in the total power consumption of the TM system.

### B. ANN ARCHITECTURE AND TRAINING

For $T_b$ control, the TM system is divided into three subsystems with outputs of $m_f$, $T_{bi}$, and $T_b$, respectively. An ANN is then implemented to model each subsystem. For the ANN-based modeling, we adopt a nonlinear auto-regressive network with exogenous inputs (NARX), a type of dynamically driven recurrent ANN [20], [21], considering the time-series vehicle data and network complexity. The ANN inputs include $\mathbf{X_C}$, $\mathbf{X_E}$, and $\mathbf{X_F}$. The time-delayed values of $\mathbf{X_C}$, $\mathbf{X_E}$, and $\mathbf{X_F}$ are also used as the inputs, as shown in Fig. 2, improving the modeling accuracy. The maximum time delays for the inputs are set to $d_c$, $d_e$, and $d_y$ min, respectively. The input data are normalized to values between –1 and 1 via a pre-processor to facilitate the training process [22]. A post-processor is then required to reverse-transform the normalized output data $y^t$ into the same unit as the original data $Y^t$.

The training is performed with open feedback loops, so that the actual data on $\mathbf{X_F}$ can be fed into the ANNs. During the training, the weighting coefficients and biases (i.e., $IW_{ij}$, $HW_{ij}$, $LW_j$, $b_{j,h}$, and $b_o$ in Fig. 2) are determined for all input, hidden, and output neurons. Once the training is finished, ANN testing is conducted with closed feedback loops, so that the ANN-based models can be used for optimal TM system operation during a period of $N_t$ min, as discussed in Section III. The ANN modeling performance is estimated using the normalized mean squared error (NMSE) [20] as:

$$e_{NMSE} = 1 - \sqrt{\frac{\sum_{k=1}^{N_Y}(Y_k - Y_k')^2}{\sum_{k=1}^{N_Y}(Y_k - Y_{k,avg})^2}}, \text{ where } Y_{k,avg} = \frac{1}{N_Y}\sum_{k=1}^{N_Y}Y_k, \quad (1)$$

where $Y_k'$ is the predicted value of $Y_k$.

### C. INTERCONNECTION OF INDIVIDUAL ANNS

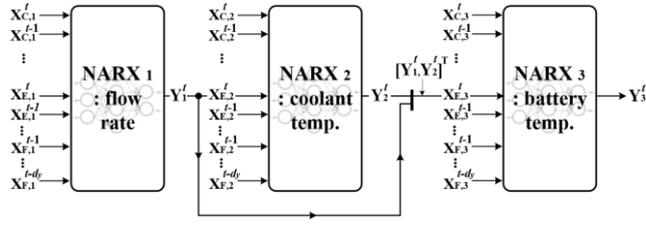

**FIGURE 3.** Interconnections of individual ANNs based on the physical configuration of the TM system.

The ANN-based models of the TM subsystems are interconnected based on the physical configuration of the TM system. As shown in Fig. 3, the output of an ANN is used as the input and corresponding time-delayed inputs of another ANN. Specifically, the output $Y_1^t = m_f^t$ of NARX$_1$ is fed into $\mathbf{X_{E,2}}^t$ of NARX$_2$ for the period from $t$ to $t-d_e$ to calculate $Y_2^t = T_{bi}^t$. Both $Y_1^t$ and $Y_2^t$ are then used for $\mathbf{X_{E,3}}^t$ of NARX$_3$ to estimate $Y_3^t = T_b^t$ during the same time period. The ANN interconnection, shown in Fig. 3, can be applied to both integrated and separate modes, discussed in Section II-A, because the inputs and outputs of the ANN-based models still include $m_f$, $T_{bi}$, and $T_b$, as shown in Table 3. The outputs of NARX$_{1-3}$ remain the same in both modes. Moreover, each of NARX$_1$ and NARX$_3$ has the same inputs for both modes and NARX$_2$ still requires $m_f$ in $\mathbf{X_E}$. Note that the training of the ANN-based TM system model, shown in Fig. 3, is not required, because it is implemented using the trained ANNs for the TM subsystems, discussed in Section II-B. The testing performance has been verified using $e_{NMSE}$ (see (1)), as described in Section IV-B.

**TABLE 3.** Controllable inputs, environmental inputs, and outputs of the individual ANNs for the TM subsystems

| Variables | | Individual ANNs | | |
|---|---|---|---|---|
| | | NARX$_1$ | NARX$_2$ | NARX$_3$ |
| **Y** | | $m_f$ | $T_{bi}$ | $T_b$ |
| Integrated | $\mathbf{X_C}$ | $P_p$ | $P_{fn}$ | - |
| | $\mathbf{X_E}$ | $t$ | $t, m_f, T_{ab}, V_{vs}, T_{pe}$ | $t, T_{bi}, m_f, I_b$ |
| Separate | $\mathbf{X_C}$ | $P_p$ | $P_{fn}, P_{cp}$ | - |
| | $\mathbf{X_E}$ | $t$ | $t, m_f, T_{ab}, I_b, AC$ | $t, T_{bi}, m_f, I_b$ |

### D. ANN SELECTION FOR LEAST OVER-FITTING

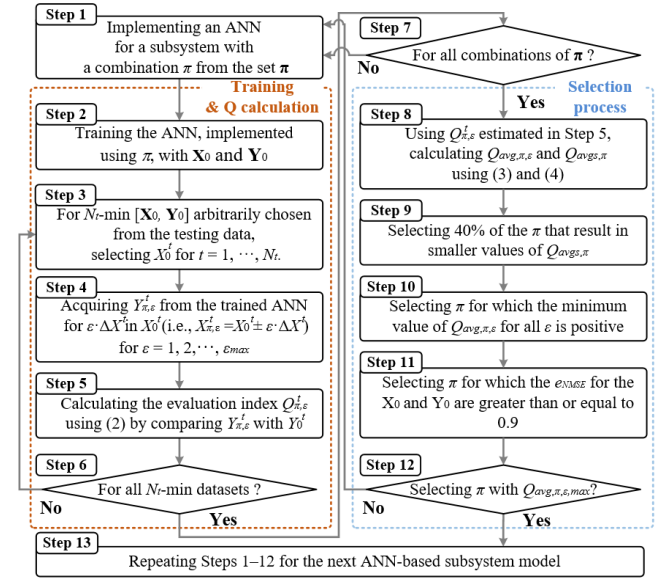

**FIGURE 4.** Flowchart of the procedure used to select the appropriate ANN architecture for the least over-fitting.

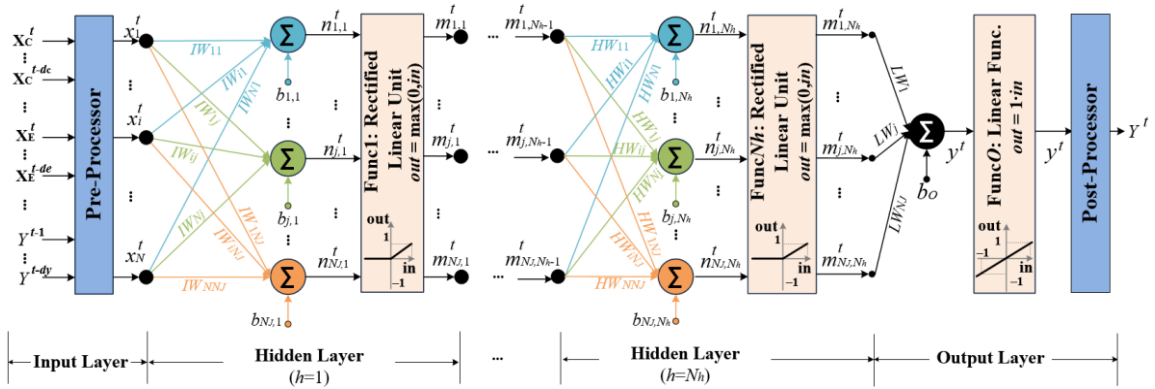

**FIGURE 2.** ANN-based model of the TM subsystem used to estimate the outputs $\mathbf{Y}^t$ for the controllable, environmental, and feedback inputs.

Data volume and variability affect the training performance of ANNs [23], implying the risk of over-fitting ANNs to limited sets of training data. Recently, various techniques [24], [25] have been discussed to mitigate the over-fitting, such as dropouts, ensemble learning, cross-validation, and early-stopping training. In this paper, a rather simple procedure is adopted, as shown in Fig. 4, where the hyper-parameters of each ANN are selected for the least over-fitting. This enables the individual ANNs to reflect the general operating characteristics of the TM subsystems more accurately, rather than only those of the training datasets. The hyper-parameter selection scheme is conducted as follows:

(*Step* 1) Implement an ANN for a subsystem with a combination $\pi$ of the numbers of hidden layers and neurons from set $\pi$, as shown in Table 4;

(*Step* 2) Train the ANN with the original input and output datasets $[\mathbf{X}_0, \mathbf{Y}_0]$;

(*Step* 3) For the $N_t$-min input and output datasets arbitrarily chosen from the testing datasets, select controllable or environmental inputs $X_0^t$ for $t = 1, \cdots, N_t$;

(*Step* 4) Acquire the outputs $Y_{\pi,\varepsilon}$ from the trained ANN for the incremental variation $\varepsilon \cdot \Delta X^t$ in the selected input $X_0^t$ (i.e., $X_{\pi,\varepsilon}^t = X_0^t \pm \varepsilon \cdot \Delta X^t$) while increasing $\varepsilon$ from 1 to $\varepsilon_{max}$;

(*Step* 5) Calculate the evaluation index $Q_{\pi,\varepsilon}^t$ for the comparison between $Y_0^t$ and $Y_{\pi,\varepsilon}^t$ as follows:

$$Q_{\pi,\varepsilon}^t = \pm (Y_{\pi,\varepsilon}^t - Y_0^t) \cdot (\Delta X^t / |\Delta X^t|), \quad \forall t, \forall \varepsilon, \forall \pi, \quad (2)$$

with the positive sign when $Y_{\pi,\varepsilon}^t$ increases, as $X_{\pi,\varepsilon}^t$ increases, and the negative sign otherwise;

(*Step* 6) Repeat *Steps* 3–5 for all the $N_t$-min datasets (in this paper, the total size of the datasets is approximately a tenth of that of the testing datasets);

(*Step* 7) Repeat *Steps* 1–6 for all the combinations of the hyper-parameter set $\pi$;

(*Step* 8) Calculate $Q_{avg,\pi,\varepsilon}$ and $Q_{avgs,\pi}$ for $t = 1, \cdots, N_t$ and $\varepsilon = 1, 2, \cdots, \varepsilon_{max}$, based on the evaluation index $Q_{\pi,\varepsilon}^t$ estimated in *Step* 5, as follows:

$$Q_{avg,\pi,\varepsilon} = \frac{1}{N_t}\sum_{t=1}^{N_t} Q_{\pi,\varepsilon}^t, \quad \forall \varepsilon, \forall \pi, \quad (3)$$

$$Q_{avgs,\pi} = \frac{1}{N_\varepsilon}\sum_{\varepsilon=1}^{N_\varepsilon}\left(\frac{1}{N_t-1}\sum_{t=1}^{N_t}\left|Q_{\pi,\varepsilon}^t - Q_{avg,\pi,\varepsilon}\right|^2\right)^{1/2}, \quad \forall \pi, \quad (4)$$

(*Step* 9) Select 40% of the combinations $\pi$ that result in smaller values of $Q_{avgs,\pi}$ than those for the rest of the combinations;

(*Step* 10) Select $\pi$ for which the minimum value of $Q_{avg,\pi,\varepsilon}$ for all $\varepsilon$ is positive;

(*Step* 11) Select $\pi$ for which $e_{NMSE}$ for the original training dataset $[\mathbf{X}_0, \mathbf{Y}_0]$ is greater than or equal to 0.9;

(*Step* 12) Select $\pi$ with the maximum value of $Q_{avg,\pi,\varepsilon}$, or enlarging the size of $\pi$ and starting from *Step* 1 for the case where no combination is left after *Step* 11;

(*Step* 13) Repeat *Steps* 1–12 for the next ANN-based subsystem model.

Using the proposed procedure, NARX$_1$ has been implemented with a relatively simple architecture including one hidden layer and 16 hidden neurons for both integrated and separate modes. This is because of the rather simple operating characteristics of the EWPs with respect to $m_f$ and $P_p$. The architecture of NARX$_2$ is characterized by two hidden layers and four hidden neurons for the integrated mode, as well as two hidden layers and six neurons for the separate mode. Similarly, NARX$_3$ has one hidden layer with 18 hidden neurons and one hidden layer with 30 hidden neurons for the integrated and separate modes, respectively. Note that different schemes to mitigate ANN over-fitting also can be integrated into the proposed SL-based strategy for optimal TM system operation.

**TABLE 4. Hyper-parameter sets for ANN-based subsystem modeling**

| Hyper-parameters $\pi$ | Integrated mode | | | Separate mode | | |
|---|---|---|---|---|---|---|
| | NARX$_1$ | NARX$_2$ | NARX$_3$ | NARX$_1$ | NARX$_2$ | NARX$_3$ |
| Number of hidden layers | {1} | {1, 2} | {1, 2} | {1} | {1, 2} | {1, 2} |
| Number of hidden neurons | {4, 6, $\cdots$, 18} | | | {4, 6, $\cdots$, 38} | | |

## III. OPTIMAL THERMAL MANAGEMENT SCHEDULING

### A. EXPLICIT REPLICATION OF TRAINED ANNS

The complete ANN-based model of the TM system, shown in Fig. 3, is represented using an explicit set of piecewise linear equations. Specifically, (5)–(8) represent the data normalization and reverse-transformation of the pre- and post-processors, respectively:

$$2X_{i,s}^t - (X_{i,s,max} - X_{i,s,min})x_{i,s}^t = \quad (5)$$
$$X_{i,s,min} + 1, \ i \in \{\mathbf{X}_{C,s}, \mathbf{X}_{F,s}\}, \quad \forall t, \forall s,$$

$$2Y_s^t - (Y_{s,max} - Y_{s,min})y_s^t = Y_{s,min} + 1, \quad \forall t, \forall s, \quad (6)$$

$$-1 \leq x_{i,s}^t \leq 1, \ i \in \{\mathbf{X}_{C,s}, \mathbf{X}_{F,s}\}, \quad \forall t, \forall s, \quad (7)$$

$$-1 \leq y_s^t \leq 1, \quad \forall t, \forall s. \quad (8)$$

Furthermore, (9) shows the linear relationship between the $i_{th}$ input neurons and the $j_{th}$ neuron in the first hidden layer:

$$n_{j,h,s}^t - \sum_{i \in \mathbf{X}_{C,s}} IW_{ij,s} x_{i,s}^t - \sum_{i \in \mathbf{X}_{F,s}} IW_{ij,s} x_{i,s}^t \quad (9)$$
$$= \sum_{i \in \mathbf{X}_{E,s}} IW_{ij,s} x_{i,s}^t + b_{j,h,s}, \quad \forall j, \forall t, \forall s, h = 1.$$

In the ANN-based model, a rectified linear unit (ReLU) [24] is used as the activation function. As shown in Fig. 5, it can be readily piecewise linearized by dividing the entire $x$-axis into two blocks (i.e., $N_{se} = 2$), requiring a single binary variable for each ReLU. Specifically, (10) indicates that the input value $n_{j,h,s}^t$ of the ReLU can be expressed as the sum of $q_{se,j,h,s}^t$, the value assigned in the segmented block $se$, for a large negative constant $r_0$. Similarly, the output value $m_{j,h,s}^t$ is equivalent to the sum of $l_{se} \cdot q_{se,j,h,s}^t$, as shown in (11). Moreover, (12)–(15) are required to set the boundaries of the segmented blocks. In (12) and (15), $v_{j,h,s}^t \in \{0, 1\}$ is used to assign a non-zero value to $q_{se=2,j,h,s}^t$ only after $q_{se=1,j,h,s}^t$ becomes equal to its maximum value (i.e., $(r_1 - r_0)$). Note that $r_1$ is zero and $r_2$ is an arbitrarily large positive constant.

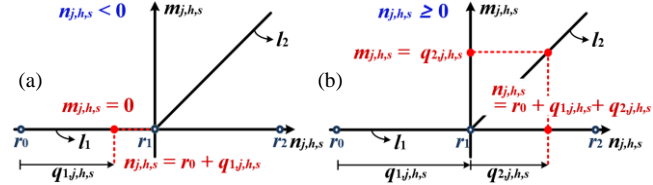

**FIGURE 5.** Piecewise linearization of a rectified linear unit (ReLU).

$$n_{j,h,s}^t - \sum_{se=1}^{N_{se}} q_{se,j,h,s}^t = r_0, \qquad \forall t, \forall j, \forall s, \forall h, \quad (10)$$

$$m_{j,h,s}^t - \sum_{se=1}^{N_{se}} l_{se} q_{se,j,h,s}^t = 0, \qquad \forall t, \forall j, \forall s, \forall h, \quad (11)$$

$$-q_{1,j,h,s}^t + (r_1 - r_0) v_{j,h,s}^t \le 0, \qquad \forall j, \forall h, \forall t, \forall s, \quad (12)$$

$$q_{1,j,h,s}^t \le (r_1 - r_0), \qquad \forall j, \forall h, \forall t, \forall s, \quad (13)$$

$$-q_{2,j,h,s}^t \le 0, \qquad \forall j, \forall h, \forall t, \forall s, \quad (14)$$

$$q_{2,j,h,s}^t - (r_2 - r_1) v_{j,h,s}^t \le 0, \qquad \forall j, \forall h, \forall t, \forall s. \quad (15)$$

As in (9), (16) shows the linear relationship between the neurons in two consecutive hidden layers for $2 \le h \le N_h$. In particular, the weighting coefficient $HW_{i,j,h-1,h,s}$ characterizes the connection from the output of the $i_{th}$ ReLU in the $(h-1)_{th}$ hidden layer to the input of the $j_{th}$ ReLU in the $h_{th}$ hidden layer, which is represented using (10). In addition, (17) represents the connections from the outputs of the ReLUs in the last hidden layer (i.e., $h = N_h$) to the normalized output $y_s^t$.

$$-\sum_{i \in N_j} HW_{ij,h-1,h,s} m_{ij,h-1,s}^t + \sum_{se=1}^{N_{se}} q_{se,j,h,s}^t$$
$$= b_{j,h,s} - r_0, \qquad \forall t, \forall j, \forall s, h = 2, ..., N_h, \quad (16)$$

$$y_s^t - \sum_{j \in N_j} LW_{j,s} m_{j,N_h,s}^t = b_{o,s}, \qquad \forall t, \forall s. \quad (17)$$

The linearized ANN model (5)–(17) can be directly integrated into the constraints of the optimization problem, as discussed in Section III-B, so that it can be readily solved using an MILP solver.

### B. OPTIMIZATION PROBLEM FORMULATION
The optimal TM scheduling is achieved by solving

$$\arg\min_{P_u^t} J_{TM} = \sum_{t=1}^{N_t} \sum_{u=1}^{N_c} c_P f_P(P_u^t) + \sum_{t=1}^{N_t} c_T f_T(T_b^t), \quad (18)$$

subject to:
- Constraints on $P_u^t$ and corresponding $T_b^t$ for the ANNs:

$$P_{u,min} \le P_u^t \le P_{u,max}, \qquad u \in \mathbf{X_C}, \forall t, \quad (19)$$

$$T_{b,min} \le T_b^t \le T_{b,max}, \qquad \forall t, \quad (20)$$

and (5)–(17),
- Constraints on the time-delayed inputs $X_{i,s}^{t-\tau}$ of the ANNs:

$$X_{i,s}^t = X_{i-1,s,pre}^{N_t}, \qquad i \in \{\mathbf{X_{C,s}}, \mathbf{X_{E,s}}\}, t = 1, \forall s, \quad (21)$$

$$X_{i-1,s}^{t-1} - X_{i,s}^t = 0, \qquad i \in \{\mathbf{X_{C,s}}, \mathbf{X_{E,s}}\}, \forall t \ge 2, \forall s, \quad (22)$$

$$X_{i,s}^t = Y_{s,pre}^{(N_t+t-(i-N_c d_c))}, \qquad i \in \mathbf{X_{F,s}}, t \le i - N_c d_c, \forall s, \quad (23)$$

$$X_{i,s}^t - Y_s^{t-(i-N_c d_c)} = 0, \qquad (24)$$
$$i \in \mathbf{X_{F,s}}, \forall t \ge i - N_c d_c + 1, \forall s,$$

- Constraints on the connections between the ANNs:

$$X_{i,s}^t - Y_{s'}^t = 0, \qquad i \in \mathbf{X_{E,s}}, \forall t, \quad (25)$$
$$(s,s') \in \{(2,1),(3,1),(3,2)\}.$$

The objective function (18) aims to minimize the total operating cost of the TM system, consisting of two terms. The first term indicates the $N_t$-min sum of the costs for the time-varying power inputs $P_u^t$ of all the controllable devices: i.e., the EWPs, fan, and compressor. The second term represents the $N_t$-min sum of the costs incurred due to the variation in $T_b^t$. Fig. 6(a) and (b) show the cost-power function $f_P$ and the cost-temperature function $f_T$, respectively. In particular, $f_T$ is implemented based on [13], [15], and [16], with slight modifications, as:

$$f_T(T_b^t) = \alpha_4 (T_b^t)^4 - \alpha_3 (T_b^t)^3 + \alpha_2 (T_b^t)^2 - \alpha_1 (T_b^t)^1 + \alpha_0, \quad (26)$$

where $\alpha_{0-4}$ are set to 0.0001, 0.0026, 0.0310, 0.6210, and 15.0681, respectively. Note that for the application of an MILP solver, $f_T$ is piecewise linearized, as shown in Fig. 6(b). Moreover, $c_P$ and $c_T$ are used to impose the relative weights on the costs for the first and second terms, respectively.

Furthermore, the constraints (19) and (20) specify the limits of $P_u^t$ and $T_b^t$, respectively, for all $u$ and $t$, ensuring the reliable operation of the TM system. The upper and lower limits can be determined based on the specifications of the EV battery and the controllable devices (i.e., the EWPs, fan, and compressor). Note that $T_b^t$ is estimated for $P_u^t$ using the linearized ANN-based model (5)–(17) of the TM system.

In addition, (21)–(24) represent the constraints on the input neurons of the ANNs for the TM subsystems that receive the time-delayed data on $\mathbf{X_C}$, $\mathbf{X_E}$, and $\mathbf{X_F}$. Specifically, (21) and (23) show that for the initial time period of the current scheduling, the input neurons receive the data on the optimal $\mathbf{X_C}$ and $\mathbf{Y}$ that were determined for $\mathbf{X_E}$ during the previous scheduling time period. Constraints (22) and (24) are then applied to the input neurons for the rest of the current scheduling time period. The $N_t$-min-ahead forecasted data on $\mathbf{X_E}$ and corresponding time-delayed data are also fed into the input neurons of the ANNs, as shown in (21) and (22).

Moreover, (25) represents the connections between the ANN-based subsystem models, as discussed in Section II-C. Specifically, the ANN output for the $s'_{th}$ subsystem is fed into the input neuron for $\mathbf{X_E}$ of the ANN for the $s_{th}$ subsystem. The output is also delivered to the input neurons for the time-delayed data on $\mathbf{X_E}$ via (21) and (23). Note that the set of ($s$, $s'$) can be pre-determined using the physical configuration of the TM system, as discussed in Section II-B.

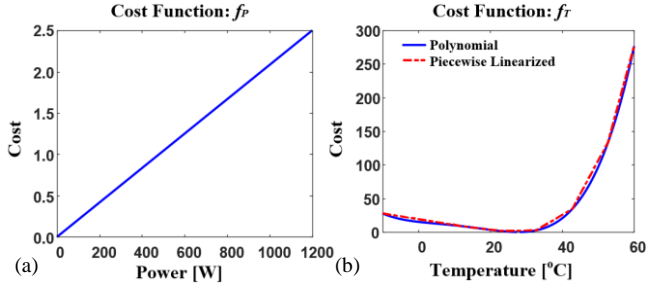

**FIGURE 6.** Cost functions for (a) the power inputs of the controllable devices and (b) the battery temperature.

## IV. CASE STUDIES AND RESULTS
### A. TEST CONDITIONS

**TABLE 5.** Main specifications of the test TM system for case studies

| Devices | Descriptions | Parameters | Values |
|---|---|---|---|
| EWP | rated power input | $P_{p, max}$ [W] | 50 |
|  | minimum power input | $P_{p, min}$ [W] | 3.4 |
| Fan | rated power input | $P_{fn, max}$ [W] | 250 |
|  | minimum power input | $P_{fn, min}$ [W] | 0 |
| Compressor | rated power input | $P_{cp, max}$ [W] | 4500 |
|  | minimum power input | $P_{cp, min}$ [W] | 0 |
| Battery | maximum temperature | $T_{b, max}$ [°C] | 48 |
|  | minimum temperature | $T_{b, min}$ [°C] | 8 |

The proposed SL-based strategy was tested for a TM system, characterized by the main specifications and operating data, shown in Table 5 and Fig. 7, respectively. Specifically, the rated power inputs of the EWP, fan, and compressor were specified as 50 W, 250 W, and 4500 W, respectively, based on the operating data. The minimum power input of the EWP was determined to be 3.4 W for the continuous circulation of the coolant. Moreover, the minimum and maximum temperatures of the battery module were set to 8°C and 48°C, respectively, under normal driving conditions. Note that as shown in Table 5, the proposed strategy does not require the physics-based modeling parameters of the TM system, unlike the conventional strategies listed in Table 1.

Fig. 7 shows the operating data of the TM system for the repeated driving cycles consisting of the UDDS and HWFET schedule. The operating data were sampled every 5 s and then averaged over every 60 s to generate the training and testing datasets for the ANNs. Consequently, the optimal TM scheduling was achieved with a sampling time of 1 min. Specifically, Fig. 7(a)–(c) show the 10-min profiles of $\mathbf{X}_C^t = [P_p^t, P_{fn}^t, P_{cp}^t]$. Fig. 7(d)–(f) represent the corresponding outputs $\mathbf{Y}_C^t = [m_f^t, T_{bi}^t, T_b^t]$ under the environmental conditions that were characterized mainly by $\mathbf{X}_E^t = [t, V_{vs}^t, I_b^t, T_{pe}^t, AC^t]$, shown in Fig. 7(g)–(j). Note that $T_{pe}^t$ and $AC^t$ affect the TM system operation for the integrated and separate modes, respectively. The variations in $P_{fn}^t$ and $P_{cp}^t$ were relatively large and continuous, whereas those in $P_p^t$ and $m_f^t$ were rather small and discretized. Moreover, for all the initial values of $T_b^{t=0}$, the variations in $T_b^t$ were small and continuous, mainly due to the large thermal capacity of the battery. Note that in several profiles, $T_{bi}^t$ changed with relatively large variations. It also can be seen that for the EV, $I_b^t$ was mainly affected by $V_{vs}^t$, resulting in the similarity between the profiles of $I_b^t$ and $V_{vs}^t$.

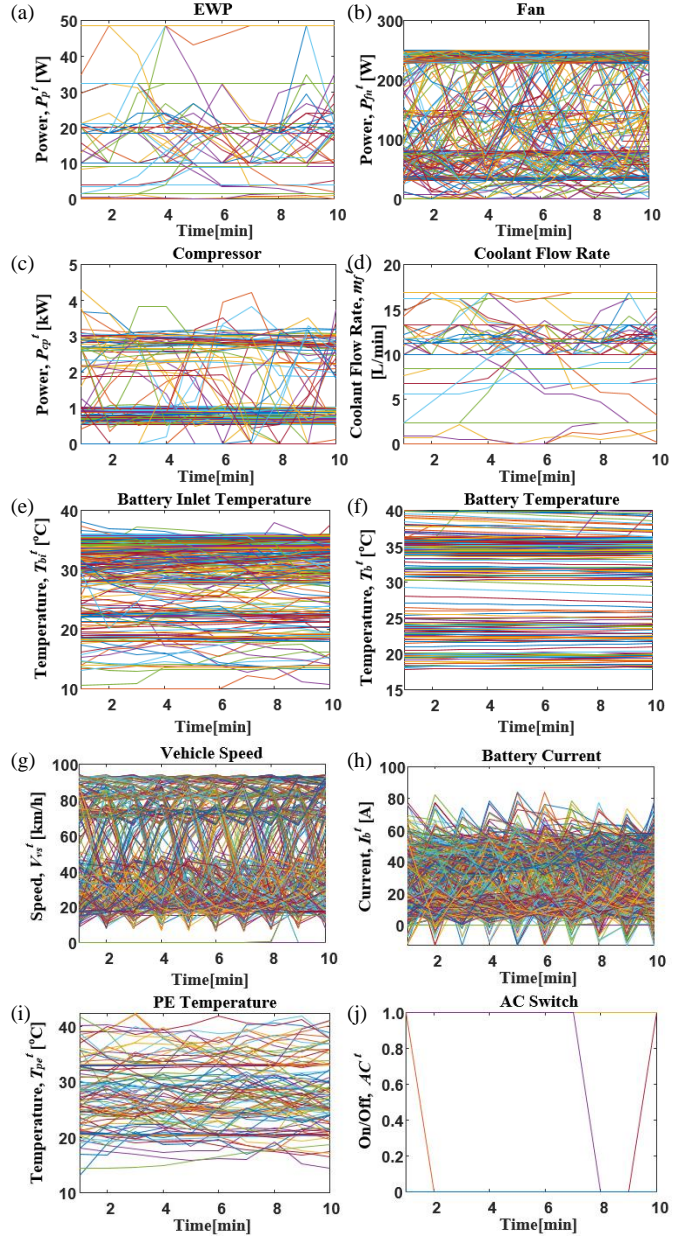

**FIGURE 7.** Operating data of the test TM system: (a)–(c) power inputs of the EWP, fan, and compressor, (d)–(f) coolant flow rate, coolant temperature, and battery temperature, (g)–(j) vehicle speed, battery current, PE module temperature, and AC switch operating status.

The total number of historical data $[\mathbf{X}_C^t, \mathbf{X}_E^t, \mathbf{Y}_C^t]$ was estimated to be 1009 and 10 with respect to time and objects, respectively, for the integrated mode. For the separate mode, it was estimated to be 2959 and 11 with respect to time and objects, respectively. Note that the time-delayed data for the objects were not considered in the estimation. The ANN training was conducted using approximately 80% and 90% of the total datasets for the integrated and separate modes, respectively. For the separate mode, the TM system operates with more controllable inputs, requiring more training data. The remaining datasets were used to test the performance of the ANN-based TM system model.

In addition, Fig. 8 shows the 30-min profiles of $T_{pe}^t$ and $I_b^t$ for $V_{vs}^t$, corresponding to the UDDS followed by the HWFET schedule, so as to test the proposed SL-based strategy for optimal TM scheduling. For simplicity, it was assumed that the ambient temperature was constant at 20°C and 40°C for the integrated and separate modes, respectively [5], during the scheduling time period. For the 30-min driving profile (i.e., $t_{dr,max}$ = 30 min), $N_t$ was set to 10 min to reflect the time-varying conditions during EV driving more accurately. Therefore, the proposed scheduling strategy was iteratively performed with respect to time. The performance of the proposed strategy was then evaluated in comparison with that of a conventional, rule-based strategy [11], [26], where the EWPs, fan, and compressor operate based on the relative differences between the temperatures of the battery, coolant, and refrigerant. The rule-based strategy has been widely applied in practice, because it requires no physics-based modeling parameters, as in the proposed strategy.

For the proposed strategy, the optimization problem (5)–(25) was formulated using MATLAB and solved with the library function of IBM ILOG CPLEX Optimizer [27]. The case studies were conducted on a computer with 32 GB of memory, an Intel Core i7-7600k @ 3.80GHz CPU, and an NVIDIA GeForce GTX 1060 3GB GPU.

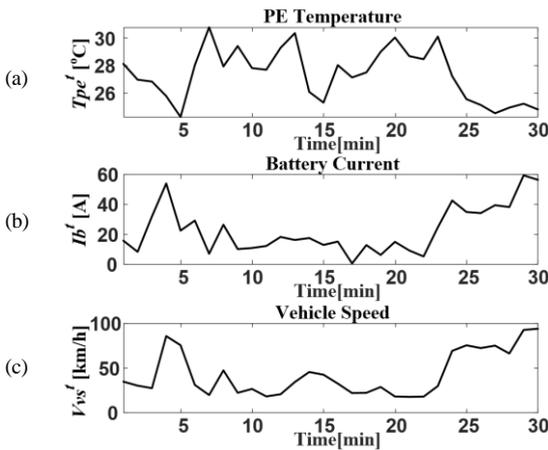

FIGURE 8. Driving profiles for optimal scheduling of the TM system operation: (a) PE temperature, (b) battery current, (c) vehicle speed.

### B. ANN TRAINING AND TEST RESULTS

The individual ANNs for the TM subsystems were trained separately using the MATLAB function [28]. Fig. 9(a)–(c) show the training results of the three NARXs with the outputs of $m_f$, $T_{bi}$, and $T_b$, respectively. The $x$- and $y$-axes represent the actual and predicted values, respectively, and the blue and green "×" marks represent the results for the integrated and separate modes, respectively. Each NARX successfully reflected the operation of each TM subsystem, leading $e_{NMSE}$ to be close to one, as shown in Table 6.

In addition, Fig. 9(d)–(f) show the testing results of the trained ANNs for the predictions on $m_f$, $T_{bi}$, and $T_b$, respectively. The test was performed after interconnecting the ANNs, as shown in Fig. 3. Table 6 also confirms the consistency between the actual and predicted data, indicating that the interconnected ANN successfully reflected the operating characteristics of the TM system for both the integrated and separate modes. This also ensures the feasibility and reliability of the optimal scheduling results, as discussed in Section IV-C. In particular, the ANN-based TM system model successfully reflected the variations in $m_f$ during the EV driving, as shown in Fig. 9(d), although the variability of the training data on $m_f$ was rather limited. This verifies that the hyper-parameter selection scheme, shown in Fig. 4, is effective in mitigating the over-fitting. Fig. 9(e) shows that the $T_{bi}$ prediction performance was rather degraded particularly for the integrated mode. However, it did not significantly affect $T_b$, as shown in Fig. 9(f), due to the large thermal capacity of the battery. Fig. 9(e) and (f) show that $T_{bi}$ and $T_b$ remained higher in the separate mode than in the integrated mode, as discussed in Section II-A.

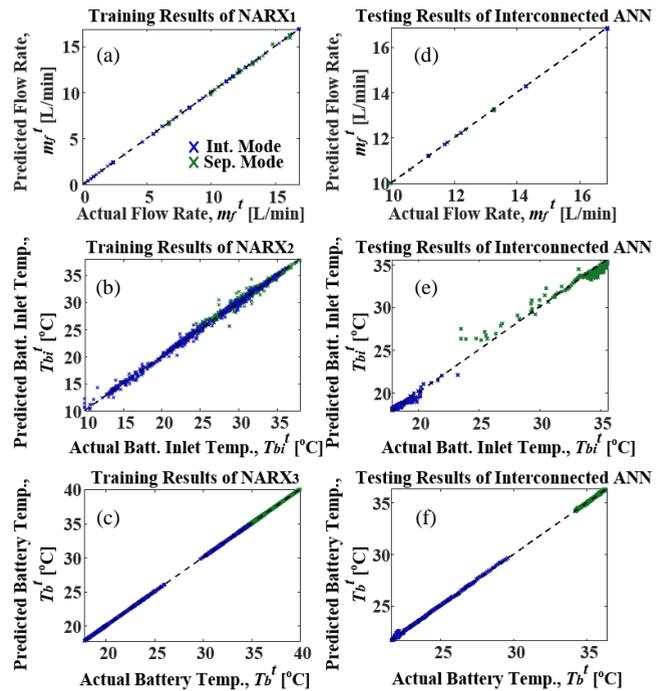

FIGURE 9. Results for (a)–(c) the training of the individual ANNs and (d)–(f) the testing of the interconnected ANN.

TABLE 6. NMSEs for the individual ANNs and interconnected ANN

| $e_{NMSE}$ | Integrated mode | | | Separate mode | | |
|---|---|---|---|---|---|---|
| | NARX$_1$ | NARX$_2$ | NARX$_3$ | NARX$_1$ | NARX$_2$ | NARX$_3$ |
| Training | 1.000 | 0.996 | 1.000 | 0.999 | 0.990 | 0.999 |
| Testing | 0.999 | 0.904 | 0.998 | 1.000 | 0.918 | 0.995 |

## C. OPTIMAL SCHEDULING RESULTS

### 1. INTEGRATED MODE

Fig. 10 shows the optimal schedule of the TM system operation for the proposed strategy (red line) in comparison with the rule-based schedule for the conventional strategy (blue line). Specifically, Fig. 10(a) and (b) present the profiles of $P_p^t$ and $P_{fn}^t$, respectively, and Fig. 10(c) and (d) depict the corresponding $T_{bi}^t$ and $T_b^t$, respectively. Note that $P_p^t$ for the PE- and battery-side EWPs are the same for the integrated mode.

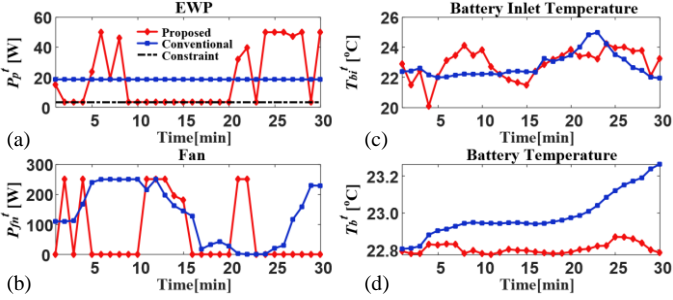

**FIGURE 10.** Optimal schedule for the integrated mode: (a), (b) power inputs of the PE-side EWP and fan, (c) coolant temperature, and (d) battery temperature.

**TABLE 7.** Comparisons of the total energy consumptions for the proposed and conventional strategies in the integrated mode

| Strategies | Total energy consumption, $E_{ec}$ [Wh] | | | |
|---|---|---|---|---|
| | Two EWPs | Fan | Compressor | Total |
| Conventional | 19.30 | 104.43 | - | 123.73 |
| Proposed | 13.62 | 50.13 | - | 63.75 |

In the proposed strategy, the fan was mainly exploited to reduce $T_{bi}^t$, mitigating a rise in $T_b^t$ due to $I_b^t$, for 1 min $\leq t_{dr} \leq$ 4 min and 11 min $\leq t_{dr} \leq$ 15 min. For 4 min $\leq t_{dr} \leq$ 10 min, the EWP increased $P_p^t$ and hence $m_f^t$ to improve the heat exchange between the coolant and ambient air at the LTR, so that $T_b^t$ remained almost constant. As $V_{vs}^t$ started increasing from $t_{dr} =$ 21 min, both EWP and fan operated to prevent an abrupt increase in $T_{bi}^t$ particularly for 21 min $\leq t_{dr} \leq$ 22 min. For $t_{dr} \geq$ 24 min, $P_p^t$ increased to almost $P_{p,max}$, while $P_{fn}^t$ remained at zero. This is because as $V_{vs}^t$ increased, the intake air flow rate at the LTR increased, so the fan did not need to operate. Moreover, the EWP operated with the maximum levels of $P_p^t$ and $m_f^t$ to exploit more effectively the increased rate of the heat exchange at the LTR, which resulted from the increased air flow rate. This feature enabled the considerable reduction of the total energy consumption $E_{ec}$, while still ensuring the control of $T_b$ at lower levels, compared to the conventional, rule-based strategy. Note that in the conventional strategy, $P_p^t$ remained constant and, therefore, the control of $T_b^t$ heavily relied on the operation of the fan. Table 7 shows that for the proposed strategy, $E_{ec}$ was reduced by 48.5%, compared to $E_{ec}$ for the conventional strategy.

### 2. SEPARATE MODE

Analogously, Fig. 11 shows the optimal operating schedule and rule-based schedule for the proposed and conventional strategies, respectively, when applied to the separate mode.

Fig. 11(a)–(c) represent the profiles of $P_p^t$, $P_{fn}^t$, and $P_{cp}^t$, respectively, and Fig. 11(d)–(e) show the corresponding $m_f^t$, $T_{bi}^t$, and $T_b^t$, respectively.

For the proposed strategy, the compressor operated with almost the maximum power input for 1 min $\leq t_{dr} \leq$ 5 min, due to the relatively high $T_{bi}^t$ and $T_b^t$ (see Fig. 11(c), (e), and (f)). Consequently, $T_{bi}^t$ and $T_b^t$ were reduced for 5 min $\leq t_{dr} \leq$ 15 min. The time delays were attributed to the large thermal capacity and hence slow time response of the battery. For $t_{dr} \geq$ 10 min, $P_{cp}^t$ started changing dynamically and rather periodically with a period of approximately 5 min. The battery-side EWP was then controlled in coordination with the compressor, as shown in Fig. 11(a) and (d), which enabled the cooling rate supplied by the compressor to be delivered to the battery more effectively. In other words, the proposed strategy induced the pre-cooling and dynamic operation of the TM system, reducing $E_{ec}$ by 6.9%, compared to the conventional strategy (see Table 8). Moreover, $T_b^t$ remained at lower levels with the proposed strategy than with the conventional strategy. Note that for the separate mode, the compressor operation increased $E_{ec}$ significantly and led to the large reduction of $T_b^t$ in both the proposed and conventional strategies.

For $t_{dr} \geq$ 25 min, the proposed strategy led to higher energy consumption than the conventional strategy, in which case the compressor turned off and the fan and EWP operated with almost constant, low power inputs. In other words, the ANNs could not reflect all of the operating conditions of the TM system, given the training datasets shown in Fig. 7. This implies that the performance of the proposed strategy can be further improved via the integration with online SL [29], [30], which is left for future work.

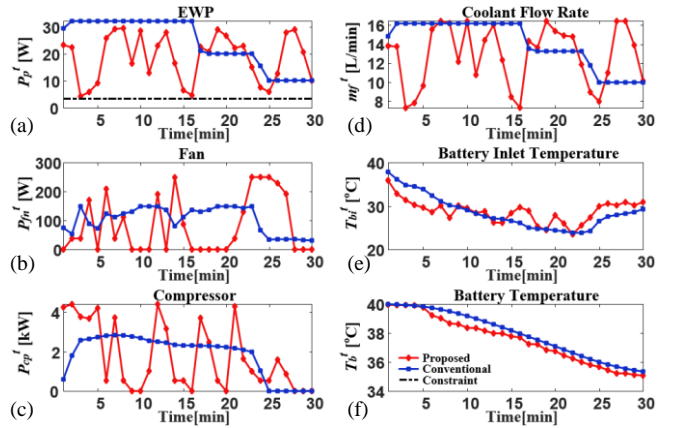

**FIGURE 11.** Optimal schedule for the separate mode: (a)–(c) power inputs of the battery-side EWP, fan, and compressor, (d), (e) coolant flow rate and temperature, and (f) battery temperature.

**TABLE 8.** Comparisons of the total energy consumption for the proposed and conventional strategies in the separate mode

| Strategies | Total energy consumption, $E_{ec}$ [Wh] | | | |
|---|---|---|---|---|
| | Single EWP | Fan | Compressor | Total |
| Conventional | 12.21 | 51.99 | 920.10 | 984.30 |
| Proposed | 9.35 | 41.19 | 865.80 | 916.34 |

## V. CONCLUSIONS

The main contribution of this paper is the development of a new SL-based strategy for optimal TM system operation, in which the total energy consumption of the EWPs, fan, and compressor is minimized while maintaining the battery temperature within an acceptable range. The main features of the proposed SL-based strategy are summarized as follows:

- An ANN was implemented and trained for each TM subsystem. The individual ANNs were then interconnected based on the physical relationships between the inputs and outputs of the subsystems, improving the accuracy of the ANN-based TM system model.
- The ANN training was integrated with the scheme to select the ANN architectures for the least over-fitting, enhancing the generalization capability of the ANN-based model in reflecting the operating characteristics of the TM system.
- The ANN-based model was represented using an explicit set of piecewise linearized equations, which could be directly integrated into the optimization problem. This enabled the application of an off-the-shelf MILP solver to the problem, ensuring the global optimality of the solution within reasonable computational time.

The proposed SL-based strategy was tested on a common TM system. The case study results confirmed that the ANN-based model successfully reflected the operating characteristics of the TM system, enabling accurate estimation of the battery temperature under various conditions on the controllable and environmental inputs. Using the ANN-based model, the coordination among the EWPs, fan, and compressor was successfully achieved in the proposed SL-based strategy, reducing the total energy consumption of the TM system by 48.5% and 6.9% in the integrated and separate operating modes, respectively, compared to the conventional, rule-based strategy. The battery temperature was also regulated successfully within the acceptable ranges.

The proposed SL-based strategy relieved the necessity of obtaining and estimating numerous physics-based modeling parameters, so that it could be widely applied to various types of TM systems. Further work is still required particularly with regard to the integration of the proposed strategy with online learning. After the proposed strategy has been initiated, the ANN-based model will be continuously trained online, as new data on the TM system operations start to be collected for various driving environments. This process will further improve the performance of the ANN-based model and hence the optimal TM scheduling strategy.